\documentstyle[11pt,aaspp4]{article}
\newcommand{\etal}{{\it et al.\ }}

\newcommand{\avg}[1]{\left\langle{#1}\right\rangle}

\begin{document}

\Large 

\title{A New Class of Estimators for the \\
            $N$-point Correlations
      }
\normalsize 
\vskip 0.5cm 
\author{Istv\'an Szapudi$^1$ and Alexander S. Szalay$^2$} 
\vskip 1cm 
\affil{$^1$NASA/Fermilab Astrophysics Center, Fermi National Accelerator Laboratory, Batavia, IL 60510-0500}
\affil{$^2$ Johns Hopkins University, Baltimore, MD 21218 }
\vskip 0.3cm

\vskip 3cm 
\centerline{\bf Abstract} 

\small 
A class of improved estimators is proposed for N-point correlation
functions of galaxy clustering, and for discrete spatial random
processes in general.  
In the limit of weak clustering, the variance
of the unbiased estimator converges to the continuum value much faster
than with any alternative, all terms giving rise to a slower
convergence exactly cancel. Explicit variance formulae are provided for
both Poisson and multinomial point processes using 
techniques for spatial statistics reported by Ripley (1988).
The formalism naturally includes most previously used
statistical tools such as $N$-point correlation functions and their
Fourier counterparts, moments  of counts-in-cells, and moment correlators. 
For all these, and perhaps some other statistics our
estimator provides a straightforward means for efficient edge corrections. 

\normalsize
\vskip 0.5cm
{\bf \noindent keywords} large scale structure of the universe -- galaxies: clustering
-- methods: numerical -- methods: statistical


\section{Introduction}

As correlation functions are some of the most useful descriptors of
galaxy clustering, their accurate estimation is of utmost
importance. There is a considerable spread in opinions on what defines
an optimal estimator. Since the early work of Peebles (1980 and
references therein)  and
coworkers (e.g. \cite{pg75,fp78}) the simple $DD/RR-1$ estimator
was widely used, where $DD$ symbolically denotes
the number of galaxy pairs at a given range of separations, and $RR$
denotes the number of random pairs generated over a similar area as
the data. It has been known for some time that data close to the edges
of a survey should have different weights, and for angular
correlations an improved estimator has been introduced: $DD/DR-1$,
where $DR$ denotes pairs of random and data
points. This method computes the contributions from galaxies near the
edge in a prorated fashion, via the cross correlations (\cite{hew82}).
 Landy and Szalay (1993, hereafter \cite{ls93}) have shown that by using
$(DD-2DR+RR)/RR$ one can carry the above argument even further. In the
limit of weak clustering the variance is proportional to $1/n^2$,
i.e. Poissonian in the pair counts, whereas for all the other
estimators the leading order term is $1/n$, where $n$ is the number of
points in the survey. \cite{fkp94} employed the
Fourier equivalent of this expression, and \cite{ham93} advocated
$DD.RR/DR^2$, which, except for a small bias, behaves like the LS
estimator. \cite{berns94} has generalized the LS estimator by
including explicitly the effects from higher order clustering.
Besides of these direct estimators ensemble estimators were in use as
well.  \cite{l54,nss56,gp77} used
$\avg{(N_1-\avg{N})(N_2-\avg{N})}/\avg{N}^2$, which is essentially
equivalent to the \cite{ls93} estimator.  They did not emphasize the
difference in the variance of equivalent unbiased estimators, and to date
other forms such as $\avg{N_1 N_2}/\avg{N}^2-1$, and $\avg{N_1
N_2}/\avg{N_1}\avg{N_2}-1$ became wide spread, despite the extra variance
they contain. In the statistics
literature, probably the most comprehensive review of the pair
estimators is by \cite{ripley88}, who discusses several ways of
performing the edge corrections.

At the same time, the estimators for the higher order correlation
functions are less well understood. The first theoretical attempt in
understanding variance of higher order correlations was \cite{cbs95},
where they calculated the variance of the void probability function,
$P_0$.  Recently \cite{sc96}
identified and calculated the contributions to the variance of 
moments of counts in cells from the finite survey size,
shot noise, the geometry of the survey, as well as measurement effects
arising from the finite number of sampling cells. These theoretical
computations were applied for realistic survey properties, and the
effects of sampling were discussed in \cite{css97}, while the results
were extended for the cumulants and for the weakly non-linear regime
by \cite{scb97}.  Finally, \cite{mvh97} discussed the variance of the
bispectrum.  The importance and distinguishing power of the higher
order functions depends on how accurately they can be measured, 
thus any gain in lowering the variance is important. 

The main goals of this {\it Letter} are (i) to explain the advantageous
properties of the LS type estimators in simple terms, (ii) generalize
them for arbitrary high order, and for any statistics depending on
$N$-tuplets of discrete point processes, 
(iii) strongly advocate their use instead of traditional estimators
with larger variance. We show that for
this class of estimators the variance from shot noise is Poissonian in
the number of N-tuplets, i.e. $1/n^N$. $\S 2$ discusses the physical
reasons, why these estimators are superior to earlier ones.
In $\S 3$ a short but exact derivation of the results is provided for 
Poissonian and multinomial point processes. The latter represents a 
survey with a given number of galaxies in it. The importance
of the results, possible generalizations, and practical issues
are discussed in $\S 4$. 

\section{Minimal Variance Estimators}

According to the previous summary of estimators for second
order processes the $(DD-2 DR+RR)/RR$ estimator has
superior shot noise behavior compared to the existing
alternatives.  While the calculations in the next section
prove this statement, here a simple argument is given.

The usual $\hat\xi= DD/RR-1$ estimator can
be expressed by the sample average of $\delta$, the dimensionless
overdensity, as
  \begin{equation}
	\hat\xi = \avg{(1+\delta_1)(1+\delta_2)}_s-1 = 
		  \avg{\delta_1 + \delta_2 + \delta_1\delta_2}_s,
  \end{equation}
where $\avg{\ldots}_s$ denotes the sample average. 
This is an unbiased estimator, since in the ensemble
average $\avg{\delta}=0$,  however, the presence of the linear
$\delta$ terms will add to the variance. Using
only the absolutely necessary term, $\delta_1 \delta_2$, will obviously 
decrease the variance. The LS estimator on the other hand can be rewritten as
$\hat\xi_2=(D_1-R_1)(D_2-R_2)/RR = \avg{\delta_1 \delta_2}_s$, i.e. it
has the simplest possible structure in $\delta$. 
This estimator and its Fourier counterpart give the smallest variance, 
since any other estimator contains extra terms,
therefore the total variance can be expressed, as the variance of the
LS estimator, plus extra terms,
  \begin{equation}
	{\rm Var}\,\hat\xi = \avg{(\delta_1+\delta_2)^2 + 
	2 \delta_1^2 \delta_2 + 2 \delta_1 \delta_2^2}_s + 
	{\rm Var}( \delta_1\delta_2 ).
  \end{equation}
Although here we do not attempt to prove, that the extra term is indeed
positive, the next section shows mathematically that the LS estimator
has minimal shot noise behavior.

\cite{ripley88} has discussed extensively the Poisson variance of second
order point processes. He has shown, that the term proportional to $1/n$
in the variance of the simple estimator $\hat\xi_0$, $\hat K_0$ in
Ripley's notation, is also proportional to $u$, the perimeter for a
two dimensional survey. This implies that the effect is due
to inadequate edge-corrections, in agreement with Hewett's (1982)
original suggestion. The subtraction of the appropriate $DR$ terms is
equivalent to an optimal edge-correction.

The effect of the unnecessary terms in the estimator on the variance
is even more pronounced for the higher order functions, since there
will be a lot more terms arising through various combinatorial
expressions. Here we propose, that the obvious generalization for
higher order correlations is to create the higher order equivalents of
the LS estimator, corresponding to $\avg{\delta_1...\delta_N}$.  In
the symbolic notation, this estimator can be written as
  \begin{equation} 
	\hat\xi_N = {(D_1-R_1).(D_2-R_2)...(D_N-R_N)}/R_1..R_N,
        \label{eq:est}
  \end{equation} 
with the exact meaning discussed in the next section in rigorous
mathematical fashion.
This estimator differs from the classic estimator proposed by e.g.
\cite{pg75} for the 3-point function $(DDD-DDR)/RRR+2$ which contains
extra variance for the above explained reasons. Note, however, 
that the counts-in-cells estimators proposed by \cite{p75} for
the three point function and by \cite{p80} for the four-point function
are the counts-in-cells equivalent of the above equation. It is worth
to emphasize again, that for both counts in cells and direct
estimators, the above form eliminates the excess variance. On the other
hand this form requires some correction to obtain the irreducible,
or ``connected'' correlations. These generalizations will be discussed
in the last section. 

It will be shown next in a general fashion that the above form 
constitutes an optimal estimator in
the Poisson and multinomial limits.
The elegant formalism outlined in \cite{ripley88} enabled us to perform most
of the calculations in a compact and general form. 
This work is highly recommended as a reference for mathematical
details omitted here.

\section{Derivation of the Edge Corrected Estimator}

Let $D$ be a catalog of data points to be analyzed, and $R$ randomly
generated over the same area, with averages $\lambda$, and $\rho$
respectively.  The role of $R$ is to perform a Monte Carlo integration
compensating for edge effects, therefore eventually the limit $\rho
\rightarrow \infty$ will be taken. $\lambda$ on the other hand is
assumed to be externally estimated with arbitrary precision. 
Many interesting statistics, such as the
$N$-point correlation functions and their Fourier analogs, can be
formulated as functions over $N$-points from the catalog. The
covariance of a pair of such estimators will be calculated in the
Poisson and multinomial limits.  They correspond to the cases, where
the number of detected objects is varied or fixed {\it a priori}.  Finally
the results are given for the general case, where correlations are
non-negligible. 

Let us define symbolically an estimator $D^pR^q$, with $p+q = N$ for a 
function $\Phi$ symmetric in its arguments
  \begin{equation}
	D^qR^p = \sum\Phi(x_1,\ldots,x_p,y_1,\ldots,y_q),
  \end{equation}
with $x_i \ne x_j \in D, y_i \ne y_j \in R$. For example for the
two point correlation function $\Phi(x,y) = [x,y \in D, r \le d(x,y) \le
r+dr]$, where $d(x,y)$ is the distance between the two points, and
$[condition]$ equals $1$ when $condition$ holds, $0$ otherwise.
Ensemble averages can be estimated via factorial moment
measures, $\nu_s$ (\cite{dv72,ripley88}). 
In the Poisson limit $\nu_s = \lambda^s\mu_{s}$, 
where $\mu_s$ is the $s$ dimensional Lebesgue measure.

The general covariance of a pair of estimators is
  \begin{equation}
	\avg{D_a^{p_1}R_a^{q_1}D_b^{p_2}R_b^{q_2}} = 
	\sum_{i,j}{p_1 \choose i}{p_2 \choose i} i!
	{q_1 \choose j}{q_2 \choose j} j!\,
	S_{i+j}\lambda^{p_1+p_2-i}\rho^{q_1+q_2-j},
  \end{equation}
with
  \begin{equation}
	S_{k} = \int\Phi_a(x_1\ldots x_{k},y_{k+1}\ldots y_N)
              \Phi_b(x_1\ldots x_{k},z_{k+1}\ldots z_N)
              \mu_{2N-k}
  \end{equation} 

Here $a$ and $b$ denote possibly two different radial bins, or even
different statistics.  The expression describes the $i$-fold degeneracy   
in the $p_1+p_2$ data points from $D$, as well
as the $j$-fold degeneracy in the $q_1+q_2$ random points drawn from $R$. 
For each of these configurations the geometric phase-space $S_{i+j}$ is
different, and the shot noise contribution of each is appropriately
summed. The dependence of
$S_k$ on $a,b$, and $N$ is not noted for convenience, but they will be
assumed throughout the paper.  An estimator for the generalized
$N$-point correlation function is
  \begin{equation}
  	w_N = \frac{1}{S}  \sum_i 
	{N \choose i}(-)^{N-i} (\frac{D}{\lambda})^i (\frac{R}{\rho})^{N-i},
  \end{equation}
where $S = \int\Phi\mu_{N}$ (without subscript).  This definition can
be expressed as $(\hat D-\hat R)^N$, where $\hat{\,} $ means
normalization with $\lambda^N, \rho^N$ respectively.  In this symbolic
$N$th power, each factor is evaluated at a different point.  Simple
calculation in the limit of zero correlations yields $\avg{w_N} =
0$. For the same reason, the disconnected parts did not have to
be subtracted, which is an important simplification in the
calculation for $N \ge 4$. 
The covariance between two estimators can be evaluated as
\begin{equation}
  \avg{w_{a,N}w_{b,N}} = \sum_{i_1,i_2,i,j}{N \choose i_1}{N \choose i_2}
  {i_1 \choose i} {i_2 \choose i}i! {N - i_1 \choose j}{N - i_2 \choose j}j!
  \frac{S_{i+j}}{S^2}\lambda^{2N-i}\rho^{-j} (-)^{2N-i_1-i_2}.
\end{equation}
In the interesting limit, where $\rho \rightarrow \infty$ only $j = 0$ 
survives. Changing the order of summation yields
\begin{equation}
   \avg{w_{a,N}w_{b,N}}= \frac{1}{S^2}\sum_i S_i \lambda^{-i} i! f_{Ni}^2,
\end{equation}
with
\begin{equation}
  f_{Ni} = \sum_j {N \choose j}{j \choose i} (-1)^{N-j}.
\end{equation}
This latter can be identified as the coefficients of $\sum_N(xy)^N$,
therefore $f_{Ni} = \delta_{Ni}$. Since $\avg{w_N} = 0$, the final result is
\begin{equation}
   {\rm (co)Var}\, w_N = \frac{S_N N!}{S^2\lambda^N}.
\end{equation}
Note that this formula represents both variance and covariance
depending on whether in the definition of $S_N$ the implicit
indices $a$ and $b$ are equal or not. 

While in the above Poisson model the total number of galaxies in the
survey can vary, it is fixed in the multinomial model. This latter
case corresponds to surveys, that detect a certain number of galaxies,
and use that to estimate also the mean density. i.e. 
estimator becomes conditional given the number of
galaxies. This introduces some correlations compared to Poisson, 
which can be taken into account for further precision, 
especially when the number of galaxies in the
survey is relatively small. The normalization of the proper estimator
changes slightly: $\lambda^i \rightarrow (n)_i/v^i$, where $n$ is the
total number of objects in the survey. This normalization renders the
estimator unbiased, even when an external estimator for the mean is
inavailable. On the other hand, the above Poisson estimator has a slight
bias, to leading order $\propto {\cal O}( N(N+1)/2n)$, if an internal
estimator for the mean is used.
 For a multinomial process the
factorial moment measure is $(n)_N v^{-N} \mu_N$, with $v$, the volume
of the survey, and $(n)_N = n(n-1)\ldots (n-N+1)$, the $N$-th falling
factorial.  The covariance

\begin{equation}
 \avg{w_{a,N}w_{b,N}}  = \frac{1}{S^2}\sum_{i,i_1,i_2}S_i v^i
              i!{N \choose i_1}{N \choose i_2}
             {i_1 \choose i}{i_2 \choose i}
             \frac{(n)_{i_1+i_2-i}(-)^{i_1+i_2}}{(n)_{i_1}(n)_{i_2}}.
             \label{eq:mnres}
\end{equation}
After further reduction (which will be presented elsewhere:
\cite{ss97c}) the final result is
  \begin{equation}
  	{\rm (co)Var}\, w_N  = \frac{1}{S^2}
        {n \choose N}^{-1}\sum_i S_i v^i {N \choose i} (-)^{N-i}.
  \end{equation}
For $N = 2$ this coincides with \cite{ls93}, taking into account
that $S^2 = S_0 \simeq (S_2 v^2)^2$ up to the integral constraint,
which was neglected during the previous calculation.  
Again, the variance is inversely
proportional to the number of possible $N$-tuplets, i.e. Poisson.

\section{Discussions}

The meaning of the proposed estimator can be understood by
specifying the function $\Phi$, such that it is $1$ when the
$N$-tuplet satisfies a certain geometry (with a suitable bin
width), and $0$ otherwise. In this case the estimator will
yield the total (or disconnected) $N$-point correlations of the fluctuations
of the underlying field $\delta$. For $N \ge 4 $ these contain
extra terms compared to the connected correlations, but not as many as
the full correlations of the field $1+\delta$. 
However, this is the class of estimators which can be precisely
corrected for edge effects. 
The connected correlations can be found by subtracting all the 
possible partitions. 
Although this is not simple
to take into account in a general fashion it is feasible for any
concrete estimator. For example for $N = 4$ the connected part is 
$\avg{\delta_1\delta_2\delta_3\delta_4}_c = 
\avg{\delta_1\delta_2\delta_3\delta_4} - 
( \avg{\delta_1 \delta_2}\avg{\delta_3 \delta_4} + sym.)$.
The necessary subtraction
will induce extra variance compared to the original estimator, however, 
its order will still remain $\lambda^{-N}$.
Partitioning keeps powers of $\delta$ unchanged thus it does
not introduce lower order terms according to the arguments given
in $\S 2$; the constant of proportionality might change though. 
In summary, the connected version of our estimator 
for the $N$-point correlations  will still have a Poisson variance in the
Poisson limit, i.e. it is fully corrected for edge effects.

While so far our discussion focused on the $N$-point correlation
functions, the general nature of the assumed function $\Phi$ can incorporate
many of the statistics used in the literature, and beyond:
Fourier correlation functions ($N-1$ spectra), moments of counts
in cells, and factorial moment correlators are obvious candidates. 
For the sake of illustration, let us briefly mention the suitable $\Phi$
functions, which correspond to the above quantities. For the
$N-1$ spectra $\Phi = \sum_{\sigma(x_1,\ldots, x_N)} \Pi e^{i x_j k_j}$/N!,
where $\sigma$ refers to the sum over all possible permutations
of the points $x_i$ to render the function symmetric. 
Possibly, weights can be incorporated as well.
Then the definition for $N = 2$ is equivalent to \cite{fkp94} up to 
the shot noise power, which is not present in our estimator, since
overlapping indices are conveniently excluded in the definition.
Moments of counts-in-cells can be represented similarly.
The $N$-th factorial moment in a cell is estimated by
$\Phi = 1$ if a certain $N$-tuplet {\it fit} into the cell, $0$ otherwise.
For $N = 2$ this method will
estimate Ripley's $K_0$ function, or the average of the two-point
function over a cell.
Our estimator therefore will give $\avg{\delta^N}$, from which the cumulants 
$Q_N \propto \avg{\delta^N}_c $
can be straightforwardly calculated. 
 Finally, if $\Phi = 1$ for a 
set of $N+M$ points fit into a pair of cells separated by
distance $r$, and $0$ otherwise, factorial moment correlators
are estimated. The proposed estimator gives $\avg{\delta_1^N \delta_2^N}$,
from which the cumulant correlators (\cite{ss97a}) are simply
obtained as $Q_{NM} \propto \avg{\delta_1^N \delta_2^N}_c$. 
Generalization for $k$-th order joint moments is obvious.

The fact that the estimator is not
exactly connected is of little importance for the $N$-point correlations
and their Fourier counterparts. It is unlikely that in the near
future these methods can be pushed much beyond $N=4$, and the calculation
of the $N$-point function from the estimator of Equation \ref{eq:est}
is quite straightforward.
The other two statistics can be measured up to higher order:
it was demonstrated both in simulations (\cite{cbh95,bge95}) 
and galaxy catalogs
(\cite{p80,ssb92,gaz92,bouchet93,mss92,gaz94,sdes95,smn96}), 
that the moments can be extracted up to 10th order.  However, the connected
moments, that is cumulants and cumulant correlators (\cite{ss97a}), can be 
calculated simply via generating functions from our estimator.
 Since the correction
for connected moments does not introduce lower order terms in the
variance, the resulting statistics will be better corrected for
edge effects, than any previous methods based on moments of counts in cells,
which cannot be corrected for edge effects (\cite{sc96,css97}).

So far we proved rigorously that the proposed edge correction
is valid for a Poisson distribution. The calculation shows
how most of the shot noise is eliminated during this process.
The eliminated discreteness error 
is also an edge effect, and it can be corrected for. 
We propose the term ``edge-discreteness'' effect, to further refine
the classification of \cite{sc96}.
 Since previous estimators were not corrected for 
this extra shot noise-variance, we conjecture
that they fare worse even when correlations are present. Although
we did not prove this rigorously (that would require the generalization
of \cite{berns94} for higher order correlations), the arguments of $\S 2$
show that this is the case. The actual 
formulae for the errors, however, can only be applied to estimate
the discreteness contribution to the error (usually not dominant
at large scales except for small fields with deep exposures, as
HDF, etc.) since it does not
take into account the finite volume and edge effects. These are always present
in a realistic distribution and depend on the integral of the
correlation function,  and on the two-point and 
higher order correlation functions (\cite{sc96}).
Next we outline possible generalizations,
which could improve on the approximations if necessary
by taking into account the correlations when performing the ensemble averages.

As explained above, the explicit variance formulae can be applied reliably
when the finite volume and edge effects are not
dominating (\cite{sc96}), i.e. a sparsely sampled survey, where
the integral of the correlation function over the survey volume
is small as well. Fortunately, edge effects are mostly eliminated
by the proposed estimator, while finite volume effects can be estimated
by actually evaluating the correlation integral over the survey volume.
However, a more accurate calculation is possible, using 
factorial moment measures which include all the needed higher order
correlations, or possibly truncated at the appropriate order
if large scales are considered. For the
variance of the generalized $N$-point correlation function, a model
for up to $2N$ statistics is needed. Such a calculation for the
$2$-point function was performed by \cite{berns94}. Note however,
that any estimation of the variance on higher order correlation
functions is approximate, because all the 
systematics, and inaccuracy of the prior model will be geometrically amplified,
in addition to the fact that the interpretation of the
variance becomes less and less straightforward as the error
distribution deviates from the Gaussian (\cite{sc96}).
Nevertheless the calculation is possible although tedious
and the basic steps are outlined below.

For a general point process the factorial moment measures can be expressed as
$\nu_N = F_N \mu_N$, where $F_N$'s are the reducible $N$-point correlation 
functions. Once this is given, it is only a matter of simple but 
lengthy calculations to express all the ensemble averages needed,
using the fact that the scaling of the factorial moment measure with
$\lambda$ for a general (canonical) point process is identical to
the Poisson case.
The result is quite similar to Equation \ref{eq:mnres}, except
the falling factorials and the powers of the volume are replaced
simply by $\lambda^{-i}$, and $S_i$ will depend additionally
on the indices $i_1$ and $i_2$. 
Some simplification can be achieved by expressing 
$F_N = \lambda^N (1 + \xi + \ldots + \xi_3 +
\ldots + \xi_N) $ with the irreducible correlation functions, and, 
because of the assumed symmetry of the 
functions $\Phi$,  replacing by $\lambda^N\sum_i C^N_i \xi_i$.
The $C^N_i$'s are symmetry factors calculable 
from generating functions (\cite{ss93a}) or cluster expansion, and
$C_1 = C_N = 1, \xi_1 = 1$. This makes it possible to replace
the integrals $S_i^{i_1 i_2}$ with irreducible integrals. 
It would be beyond the scope of this work to quote the explicit
formula and explore its applicability. Our goal was only to
clearly show how such calculation can be trivially performed
if necessary, the rest is left for subsequent research.

This {\it Letter} proposed a set of edge corrected estimators for the
generalized $N$-point correlation functions. The estimators
were defined using the framework provided by recent
results in spatial statistics, which can handle in a uniform
way almost all previously used statistics for characterizing
higher order clustering. Thus our formalism includes among
others the $N$-point correlation functions, $N$-point Fourier
transforms (or $N-1$-spectra), moments of counts-in-cells,
and moment correlators. Explicit calculations were performed in the Poisson
and multinomial limits to show that the variance is approximately
Poisson, i.e. in both cases it is inversely proportional to the number 
of possible  $N$-tuplets. The calculation also pinpointed how previous
estimators were not even corrected for the ``edge-discreteness'' effects,
and physical arguments were given, that they would perform even
worse when correlations are present. 
We proposed ways to estimate the other important contributions,
the edge and finite volume effects, thus a combination
of numerical estimates combined with our formulae will yield 
substantially improved accuracy over previous techniques.

\acknowledgments

I.S. would like to thank L. Hui and S. Dodelson for discussions.
I.S. was supported by DOE and NASA through grant
NAG-5-2788 at Fermilab. A.S.Z. was supported by a NASA LTSA grant.

\hfill{}

\end{document}